# Asymmetric Heat Transfer with Linear Conductive Metamaterials

Yishu Su [1], Ying Li [2,3], Minghong Qi [2,3], Sebastien Guenneau [4], Huagen Li [5], Jian Xiong [1]*

1. Center for Composite Materials and Structures Harbin Institute of Technology Harbin 150001, China
2. Interdisciplinary Center for Quantum Information, State Key Laboratory of Modern Optical Instrumentation, ZJU-Hangzhou Global Scientific and Technological Innovation Center, Zhejiang University, Hangzhou 310027, China
3. International Joint Innovation Center, Key Lab. of Advanced Micro/Nano Electronic Devices & Smart Systems of Zhejiang, The Electromagnetics Academy of Zhejiang University, Zhejiang University, Haining 314400, China
4. UMI 2004 Abraham de Moivre-CNRS, Imperial College London, London SW7 2AZ, UK
5. Department of Electrical and Computer Engineering, National University of Singapore, Singapore 117583, Singapore

**Abstract**

Asymmetric heat transfer systems, often referred to as thermal diodes or thermal rectifiers, have garnered increasing interest due to their wide range of application possibilities. Most of those previous macroscopic thermal diodes either resort to nonlinear thermal conductivities with strong temperature dependence that may be quite limited by or fixed in natural materials or rely on active modulation that necessitated auxiliary energy payloads. Here, we establish a straightforward strategy of passively realizing asymmetric heat transfer with linear conductive materials. The strategy also introduces a new interrogative perspective on the design of asymmetric heat transfer utilizing nonlinear thermal conductivity, correcting the misconception that thermal rectification is impossible with separable nonlinear thermal conductivity. The nonlinear perturbation mode can be versatilely engineered to produce an effective and wide-ranging perturbation in the heat conduction, which imitates and bypasses intrinsic thermal nonlinearity constraints set by naturally occurring counterparts. Independent experimental characterizations of surface thermal radiation and thermal convection verified that the heat exchange between a graded linear thermal metamaterial and the ambient can be tailored to achieve macroscopic asymmetric heat transfer. Our work is envisaged to inspire conceptual models for heat transfer control, serving as a robust and convenient platform for advanced thermal management, thermal computation, and heat transport.

The development of asymmetric heat transfer devices, which could be treated as the counterpart of nonlinear solid-state devices regulating electrical conduction, such as diodes and transistors, has profound implications for thermal circuits, and thermal management. Recently, the ability to manipulate the phononic [1-6] and electronic [7-9] heat conduction has been demonstrated and offers a promising method of controlling heat flux. Thermal diode [1,2], thermal transistor [3], thermal memory [4] and thermal circuits [10,11] at the nanoscale have been experimentally or theoretically demonstrated. At the macroscale, given the lack of conceptual underpinnings, the design of thermal rectifier devices based on nonlinear materials is plagued by several misunderstandings, and it is also impossible to achieve asymmetric heat transfer using linear materials, asymmetric heat transfer systems thus have to be achieved by employing nonlinear materials [12-15] or active modulations [16-18]. And the thermal conductivities of natural materials often do not have strong temperature dependences, so asymmetry can only be accomplished with the aid of extra designs using shape-memory alloys [19,20] or other phase change materials [21,22]. One of the primary reasons for this is the absence of an overarching theoretical framework for the use of alternative mechanisms.

It is now conceivable to manipulate heat flow following human desires via artificial materials and structures [23-27]. For example, one can use specialized thermal conductivity arrangements to direct heat flow to achieve thermal cloaking [28-31], camouflaging [32-34], or transparency [35-37] in the field of heat conduction. Moreover, some studies have addressed heat convection [38-40], heat radiation [41], and the combination of these two phenomena at the same time [42] related to heat conduction, which has facilitated the development of thermal metamaterials. However, their experimental implementation is still challenging, particularly the independently accurate determination of the effect of surface thermal radiation during heat conduction and the efficient characterization of out-of-plane thermal convection that eliminates the influence of thermal radiation.

In a two-terminal system, it amounts to applying a coordinate transformation $x' = L - x$ to the heat conduction equation, which only affects the equation along the x-axis when heat flows into the system from the left and right sides. No matter how linear the thermal conductivity distribution is, when the system is in a steady state, the temperature distribution is always symmetrical. Here, a novel asymmetric heat transfer systems excogitation theory is introduced, which enables both a unique viewpoint and a more comprehensive understanding and scrutiny of the asymmetric heat transport processes resulting from nonlinear materials, correcting previous misconceptions that thermal rectification is impossible with separable nonlinear thermal conductivity [14], and enabling the construction of asymmetric heat transfer employing linear conductive metamaterials. By adding a nonlinear perturbation to the heat conduction equation, it is possible to get asymmetric heat transfer without using a material with nonlinear thermal conductivity. The graded linear thermal metamaterials, which could induce parameter mismatches in the nonlinearized equations, are used to break the fundamental symmetry. The macroscopic asymmetric heat transfer system is experimentally demonstrated utilizing graded thermal metamaterials with an optimal gradient distribution and tailoring surface thermal radiation in the vacuum chamber and quantitatively characterized surface thermal convection, respectively.

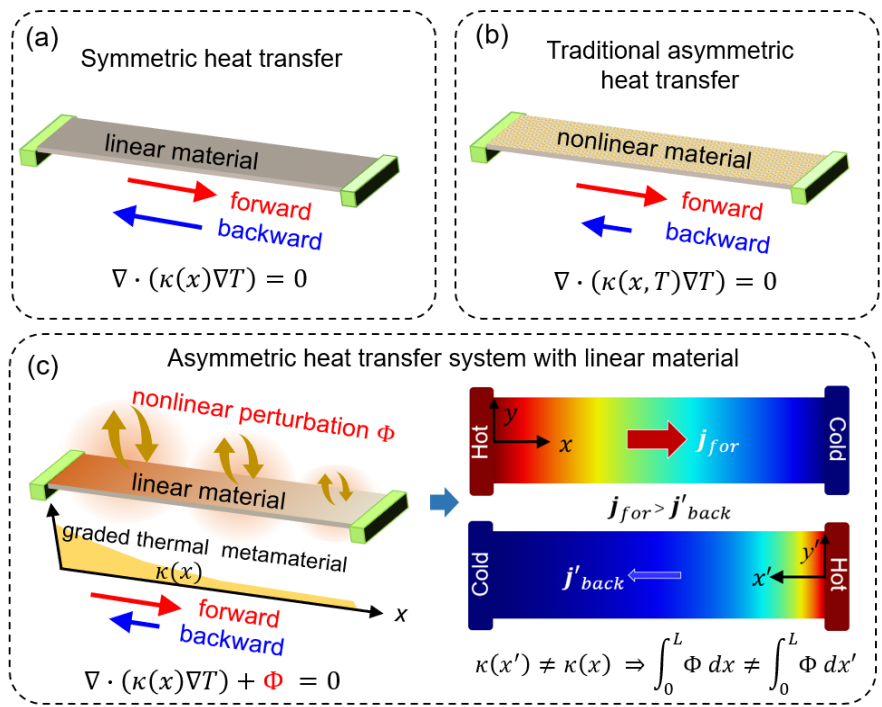

**Figure 1**. Asymmetric heat transfer system with linear conductive material. (a) The symmetric heat transfer with the linear conductive material. (b) A traditional asymmetric heat transfer system with the nonlinear conductive material. (c) An asymmetric heat transfer system with graded linear conductive material tailors the ambient perturbation inserted into the heat conduction equation and the asymmetric heat transfer temperature profile with uneven transfer heat flow caused by asymmetry thermal conductivity.

Unlike traditional asymmetric heat transfer systems (Figure. 1) in need of nonlinear material, the nonlinear perturbation $\Phi(x,T,dT/dx,d^2T/dx^2,\cdots)$ containing both the spatial element $x$ and the temperature $T$ and its derivative elements is incorporated into the heat conduction equation to investigate the potential of breaking the symmetry in the heat conduction process (Figure. 1(c)).

$$\frac{d}{dx}\left(\kappa(x)\frac{dT}{dx}\right)+\Phi(x,T,\frac{dT}{dx},\frac{d^2T}{dx^2},\cdots)=0 \tag{1}$$

The boundary condition remains $T_{left}=T_H$, $T_{right}=T_L$ ($T_H>T_L$). When applied to the heat transfer unit, the nonlinear perturbation represents the general heat flow density, which could be introduced by nonlinear thermotics [12] or be described as volumetric [40] convection or radiation, or the heat exchange with the environment, which could be surface convection [39] or radiation thermal transmission. Under the transformation $x'=L-x$ caused by the change of heat flow direction, Equation (1) can be expressed,

$$\frac{d}{dx'}\left(\kappa(L-x')\frac{dT}{dx'}\right)+\Phi(L-x',T,\frac{dT}{dx'},\frac{d^2T}{dx'^2},\cdots)=0 \tag{2}$$

The boundary condition is the same as Equation (1). In one-dimensional and isotropic media, the solution of the two equations is symmetric about $T=(T_H+T_L)/2$ when $\Phi(x,T,dT/dx,d^2T/dx^2,\cdots)=0$ in the one-dimensional case (see Supplementary Material, Note 1). When $\Phi(x,T,dT/dx,d^2T/dx^2,\cdots)\neq 0$, Equation (1) and (2) becomes nonlinear equation, it is possible to produce unequal solutions under the transformation $x'=L-x$. Considering the nonlinear perturbation $\Phi$ as the general heat flow density contributed to the heat transfer element, there will be functional inequality when thermal rectification occurs.

$$\int_0^L \Phi\left(x,T,\frac{dT}{dx},\frac{d^2T}{dx^2},\cdots\right)dx \neq \int_0^L \Phi\left(L-x',T,\frac{dT}{dx'},\frac{d^2T}{dx'^2},\cdots\right)dx' \tag{3}$$

As a sufficient necessary condition for the rectification phenomenon, the functional inequality indicates that the asymmetric heat transfer is achieved when the nonlinear perturbation term produces dissimilar feedback under the transformation $x'=L-x$. The presence of unequal solutions to Equations (1) and (2) may be inferred by comparing the forms of the two equations, which yields the conclusion that

$\kappa(L-x) \neq \kappa(x)$ (Figure. 1(c)) when the spatially-dependent term of $\Phi$ is not congruent with the variation of $\kappa$. In Equations (1) and (2), considering isotropic media, being analogous in format $\nabla \cdot (\kappa_{eff} \nabla T) = 0$, the effective thermal conductivity of the system can be represented as,

$$\kappa_{eff} = \kappa(x) + \psi(x, T, \frac{dT}{dx}, \frac{d^2T}{dx^2}, ...) \frac{dx}{dT} \tag{4}$$

The divergence of the function $\Psi(x, T, dT/dx, d^2T/dx^2, \cdots)$ is $\Phi(x, T, dT/dx, d^2T/dx^2, \cdots)$,

$$\nabla \cdot \Psi\left(x, T, \frac{dT}{dx}, \frac{d^2T}{dx^2}, ...\right) = \Phi\left(x, T, \frac{dT}{dx}, \frac{d^2T}{dx^2}, ...\right) \tag{5}$$

It can be shown from Equations (4) and (5) that the nonlinear perturbation $\Phi$ will ultimately interact with the linear material parameters to produce a possible nonlinear effect and that the nonlinear perturbation $\Phi$ is analogous to the nonlinear thermal conductivity. For example, if the nonlinear thermal conductivity has the form $\kappa(x,T) = \kappa_X(x) + \kappa_T(T)$ or $\kappa(x,T) = \kappa_X(x)\kappa_T(T)$, the equivalent nonlinear perturbation factors is $\Phi = \kappa_T \frac{d^2T}{dx^2} + \frac{d\kappa_T}{dT}\left(\frac{dT}{dx}\right)^2$ and $\Phi = \frac{\kappa_X}{\kappa_T} \frac{d\kappa_T}{dT}\left(\frac{dT}{dx}\right)^2$, respectively, and the heat conduction equation transforms to

$$\frac{d}{dx}\left(\kappa_X(x)\frac{dT}{dx}\right) + \Phi = 0 \tag{6}$$

This equation demonstrates that the asymmetric response of the accompanying nonlinear perturbation on the heat conduction equation is the essence of the nonlinear thermal conductivity that could cause asymmetric heat transfer. When $\kappa(L-x) \neq \kappa(x)$, which fulfills Equation (3), the asymmetric heat transfer with uneven heat flux (Figure. 1(c)) can be achieved under the transformation $x' = L - x$. Naturally, it is also feasible to evaluate whether the system is asymmetric heat transfer by $\kappa_X(x)$ based on this conclusion when the thermal conductivity is nonlinear in the form of $\kappa(x,T) = \kappa_X(x) + \kappa_T(T)$ or $\kappa(x,T) = \kappa_X(x)\kappa_T(T)$. The system given by Equation (6) displays the consequence of asymmetric heat transfer when $\kappa_X(L-x) \neq \kappa_X(x)$, which rectifies the misconception that resulted from disregarding the homogeneity of the functional space [14] (See Supplementary Material, Note 2 and Note 3).

To generate asymmetric heat transfer without employing nonlinear materials, the next stage is to assess realizable nonlinear perturbation $\Phi$ and fulfill the sufficient necessary conditions for linear materials. Under vacuum conditions, the convective heat transfer is suppressed, and the surface radiative heat of an inhomogeneous thermal conductivity sheet of length $L$ contributes solely to the nonlinear perturbation term $\Phi = (2L/h)\varepsilon\sigma(T_A^4 - T^4)$, where $h \ll L$ is the thickness of the sheet, $T_A$ is the ambient temperature, $\varepsilon$ is the constant emissivity of the surface and $\sigma = 5.67 \times 10^{-8}\,\text{W}/(\text{m}^2\text{K}^4)$ is the Stefan-Boltzmann constant. Equation (1) can be written as follows,

$$\frac{d}{dx}\left(\kappa(x)\frac{dT}{dx}\right) + \frac{2L}{h}\varepsilon\sigma(T_0^4 - T^4) = 0 \tag{7}$$

In Equation (7), it can be noted that the way the thermal conductivity is distributed determines whether or not asymmetric heat transfer is achievable. The solution of Equation (7) is symmetric to $x = L/2$ with the same form $T(x) = T'(L-x)$ when $\kappa(L-x) = \kappa(x)$, and this finding can also be exploited to construct the temperature trapping device [20] (See Supplementary Material Note 4). We investigated the dependence of asymmetry on how the thermal conductivity changes (See Supplementary Material, Note 5). The most obvious asymmetry occurs when the thermal conductivity exists in a linear gradient manner. In the next analysis, the linear thermal conductivity function is given the form $\kappa_{real} = kx + b$ with $k = -300$ W/m K and b = 400 W/m K.

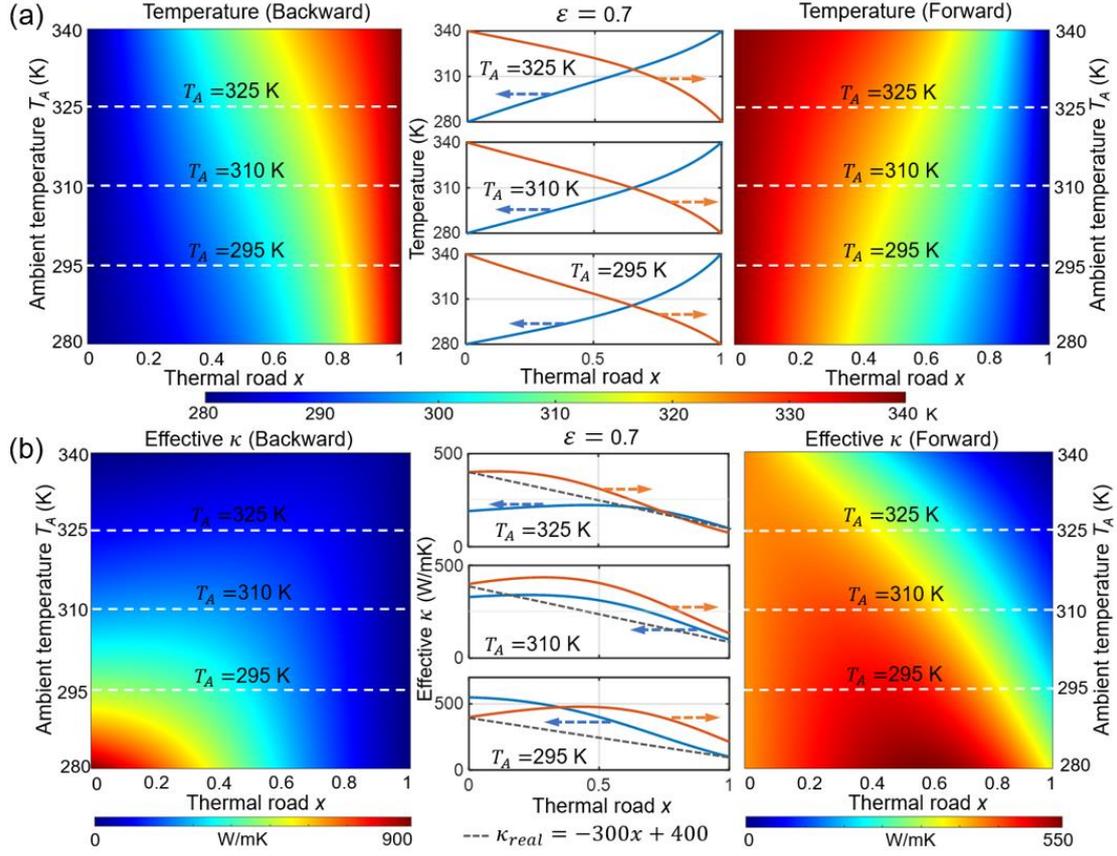

**Figure 2.** Under transformation $x' = L - x$, simulated asymmetric distribution of temperature (a) and effective thermal conductivity (b) at different ambient temperatures $T_A$ while considering the radiative heat transfer with emissivity $\varepsilon = 0.7$ as the nonlinear perturbation $\Phi$. And the actual thermal conductivity distribute with $\kappa_{real} = -300x + 400$ (W/mK, dark dashed line) when $\Phi = 0$.

Under various ambient temperatures $T_A$, the simulated asymmetric distribution of temperature and effective thermal conductivity produced by Equation (5) is shown in Figure 2. Under transformation $x' = L - x$, Equations (1) and (2) illustrate two distinct solutions while their solutions are asymmetric about $T = (T_H + T_L)/2$ when $\Phi = 0$. For example, when $T_A$ = 295 K, 310 K, or 325 K, the temperature distribution at all three temperatures is asymmetric at about $T$ = 310 K. The effective thermal conductivity $\kappa_{eff}$ distribution does not overlap along the thermal road, which indicates that, with the inclusion of the nonlinear perturbation factor, the equivalent thermal conductivity displays a temperature-interaction thermal conductivity comparable to that of the nonlinear thermal conductivity. Additionally, the effective temperature distribution is approximately symmetric due to the close value of the two effective thermal conductivity curves at $T_A$ = 310 K.

In Figure 3, the mechanism of asymmetric heat transfer is explored, and the rectification efficiency of the structure under different ambient temperatures is investigated, by calculating the heat flux of forwarding and backward thermal transfer (Figure 3a) and including the proportion of the inflow and outflow heat flow (Figure 3(b) & (c)). We calculated five distinct groups of heat fluxes in forward and backward directions at various ambient temperatures $T_A$. As seen in Figure 3(a), the heat flow distribution is asymmetrical at the same ambient temperature $T_A$. The heat inflow and outflow are not equal due to the out-of-plane heat exchange. Therefore, we individually computed the inflow and outflow rectification coefficients $R_r = |j'_{back}|/|j_{for}|$ and displayed them in Figure 3(c). It has been observed that at low ambient temperatures, inflow heat has the greatest rectification impact, whereas outflow heat flow has the greatest rectification impact at elevated temperatures. A larger thermal conductivity slope value is beneficial when it comes to rectifying heat flow. In the present model with $k = 300$, it demonstrates that the rectification coefficient of outflow heat may surpass 1.37 when the ambient temperature is 340 K, and it could approach 0.73 when the ambient temperature is 280 K.

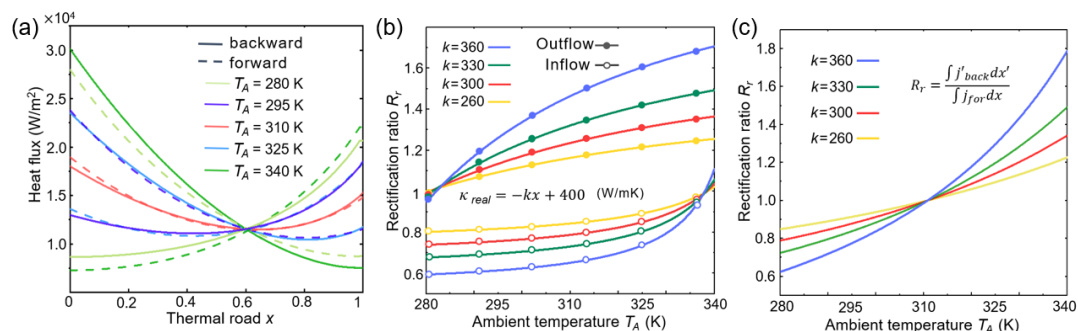

**Figure 3**. When considering the radiative heat transfer with emissivity $\varepsilon = 0.7$ as the nonlinear perturbation $\Phi$, the heat flux distribution (a) at different ambient temperatures under the forward and backward heat flows when the thermal conductivity is distributed with $\kappa_{real} = -300x+400$. (b) The rectification ratio curves of heat inflow (solid dotted line) and outflow (hollow dotted line) under various linear parameters $k$ of thermal conductivity $\kappa_{real} = -kx+400$ (W/mK). (c) The rectification ratio curves of the average heat flow under various linear parameters $k$ of thermal conductivity.

The experimental setup was schematically depicted in Figure 4(a). Using the theory of neutral inclusions, the graded thermal metamaterial with a linear thermal conductivity $\kappa_{real}$ that changes with the slope $k$ could be manufactured by filling the base material plate with various materials following the rule: $\kappa_{real} = kx + b = \kappa_b + 3f(x)\kappa_b(\kappa_0 - \kappa_b)/(3\kappa_b + (1 - f(x))(\kappa_0 - \kappa_b))$, where $\kappa_b$ and $\kappa_0$ are the thermal conductivity of the base material and the filling material respectively, and $f$ is the volume fraction of the filling material. To construct the equivalent thermal conductivity that is linearly changing according to the value of $k = -300$ (i.e. $\kappa_{real} \in [100, 400]$ W/m K), we use copper ($\kappa_b = 400$ W/m K) as the base material and cured epoxy resin ($\kappa_0 = 0.2$ W/m K) as the infill material. The graded thermal metamaterial with radiant heat dissipation film with $\varepsilon = 0.7$ was placed in the vacuum chamber with the identical film attached to the inner walls to isolate the influence of thermal convection, leaving only the surface radiative heat exchange between the sample and the surroundings. Under a temperature difference of 30 − 40 K, two heat conduction conditions in the forward and backward directions were examined for their temperature distributions, shown in Figures 4(b) and 4(c), employing ambient temperature as the low-temperature end and the high-temperature end, respectively. The red experimental data adopts the original experimental value after transformation $x = 1 - x'$, whereas the blue adopts the value after $x = 1 - x'$, $y = T_L + T_H - T_{blue}$ transformation to service standard the asymmetric heat transfer effect. It indicates that when $\Phi = 0$, the temperature distribution curve after transformation is identical to the temperature distribution curve without transformation (black dashed line), and when it's not equal to zero, there is a clear distinction. Both source-temperature systems were tested and compared under vacuum circumstances with theoretical values. The experimental results match well with theoretical values. In the higher source-temperature system (Figure 4(b)), the asymmetric impact is enhanced because the surface radiation term takes the difference of the fourth power form.

It is very difficult to determine the thermal convection effect separately under the real challenge that the thermal radiation impact is difficult to eliminate by experimental techniques. The approximate superposition property of thermal radiation and thermal convection effects followed in the heat conduction equation is sufficient to characterize the result of asymmetric heat transfer only under convective conditions (Figure 4(e)) by measuring the results of temperature profiles under natural conditions (Figure 4(d)) at the same boundary conditions and eliminating the effect of thermal radiation. The superposition approximation could be represented as $T_c = T_s + (T_n - T_r)$, where $T_c$ means the symmetric temperature (black dashed line), $T_s$ and $T_r$ represent the temperature at convective (Figure 4(e)) and radiative (Figure 4(b)) condition respectively, $T_n$ represent the natural condition (Figure 4(d)) considering both thermal convection and radiation. The calculated results for natural air convection under pure convection circumstances with a convective coefficient of 2 W/(m² K) correspond well with the theoretical outcomes.

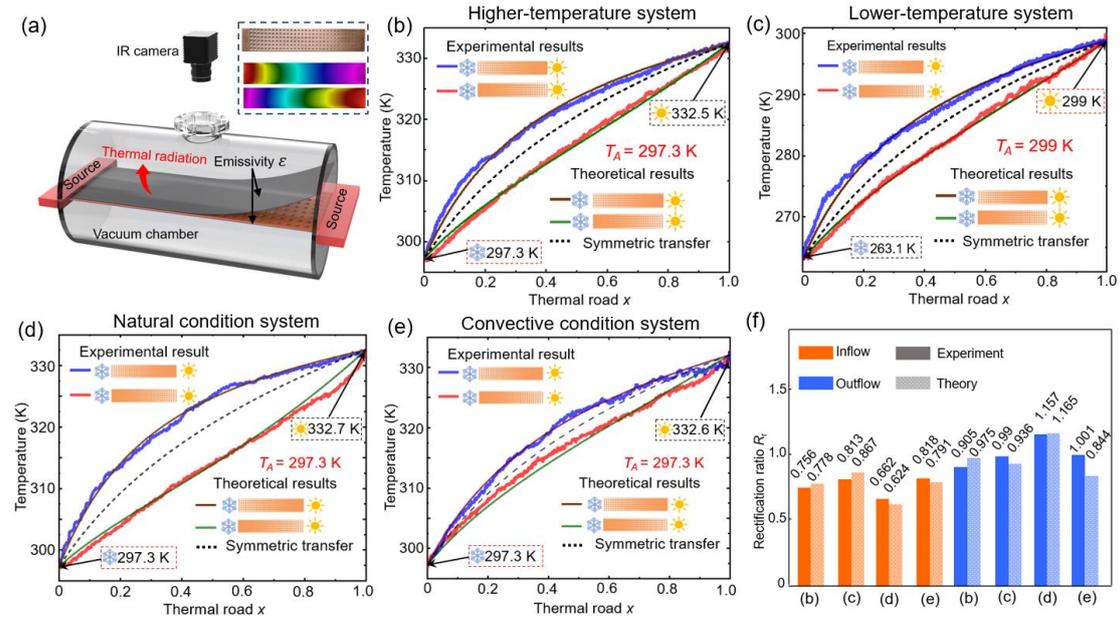

**Figure 4**. Experimental verification of asymmetric heat transfer system. (a) A radiant heat dissipation film is applied to the front and back of the sample with a particular thermal conductivity distribution and inserted in a vacuum chamber. The insert figure is the sample photograph and initial display of experimental results. Comparison between results of experiment and theory under (b) $T_L = T_A = 297.3$ K, $T_H = 332.5$ K and (c) $T_H = T_A = 299$ K, $T_L = 263.1$ K considering thermal radiation condition ($\varepsilon_{eff} \approx 0.5$) only and $k = 300$, $L = 0.15$ m, $h = 0.002$ m. The test component is seen in figures (b and c) with holes of varying sizes in a sample. The heat conductivity decreases as the hole size increases. (d) Comparison between results of experiment and theory under natural conditions considering both thermal convection and radiation. (e) The calculated result considers thermal convection only. (f) Rectification coefficient derived from experimental and theory data.

The rectification coefficients of the heat inflow and outflow (Figure 4(f)) are determined by fitting the experimental data to the thermal conductivity at both ends. Even more convincing is that the experimental and theoretical rectification coefficients are in good agreement, which shows that the experiment is accurate from an even deeper perspective. When examining solely surface radiation, the high-temperature system rectification impact is greater than that of the low-temperature system, consistent with the results in Figures 4(b) and 4(c). Due to the weak thermal convection coefficient under natural circumstances, the asymmetric heat transfer impact considering just the thermal convection effect is less than the asymmetric heat transfer effect considering only the thermal radiation effect. The nonlinear perturbation excogitation theory provides a robust platform capable of exploiting the higher thermal rectification coefficients sought by linear or nonlinear materials through bespoke nonlinear perturbation terms, combined with corresponding graded thermal metamaterials with optimal gradients. It is also feasible in further work to achieve optimal thermal rectification by the specific design of spatially-dependent surface radiation coefficients or thermal convection coefficients that match the thermal conductivity of the linear gradient distribution.

Incorporating nonlinear perturbations into the heat conduction equation provides an innovative and straightforward perspective on the asymmetric heat transfer problem in the design of nonlinear materials. That is, the key to making asymmetric heat transfer with nonlinear thermal conductivity work is to make sure the asymmetry of their spatially-independent parts is maintained. This theory offers a robust platform for the achievement of macroscopic asymmetries in heat transfer utilizing linear materials and employing graded thermal metamaterials. Proof-of-concept experiments considering surface radiation in a vacuum environment and considering both thermal radiation and convection, as well as the independent characterization of surface thermal convection using approximate superposition, prove the validity of the proposed asymmetry heat transfer. Asymmetric heat transfer phenomena can also occur in volumetric radiation or convection transmission systems that employ graded thermal metamaterials. Furthermore, when thermally related effects, such as thermoelectric effects, are considered in the heat conduction system, more asymmetric heat transfer events are possible. These discoveries provide a paradigm shift for asymmetric heat transfer, which could advance the developments of the thermal diode, thermal rectifier, and other thermal meta-devices.

**Additional information**

Supplementary information is available for this paper.

**Acknowledgments**

Y.S., M.Q., and Y.L. contributed equally to this work. The present work was supported by the Fundamental Research Funds for the Central Universities (Grant No. HIT.OCEF.2021010) and the Fundamental Research Funds for the Central Universities (2021FZZX001-19) and Young Scientist Workshop of Harbin Institute of Technology. Y.L. acknowledges the support from the National Natural Science Foundation of China (Grant No.: 92163123).

**Conflict of Interest**

The authors declare no conflict of interest.